\documentstyle[pre,multicol,aps,epsfig,subfigure]{revtex}

\begin{document}

\title{An Alternative Method to Deduce Bubble Dynamics in Single Bubble Sonoluminescence Experiments}

\author{G. Simon$^1$, and M. T. Levinsen$^2$}
       
\address{$^1$ Department of Atomic Physics, E\"otv\"os Lor\'and University,
              H-1117 Budapest, Hungary\\
         $^2$ Center for Chaos and Turbulence Studies, Niels Bohr Institute,
              Blegdamsvej 17 DK 2100, Copenhagen \O, Denmark}

\date{\today} 

\maketitle

\begin{abstract}
In this paper we present an experimental approach that allows to deduce the important dynamical parameters of single sonoluminescing bubbles (pressure amplitude, ambient radius, radius-time curve) The technique is based on a few previously confirmed theoretical assumptions and requires the knowledge of quantities such as the amplitude of the electric excitation and the phase of the flashes in the acoustic period. These quantities are easily measurable by a digital oscilloscope, avoiding the cost of expensive lasers, or ultrafast cameras of previous methods. We show the technique on a particular example and compare the results with conventional Mie scattering. We find that within the experimental uncertainties these two techniques provide similar results.

\end{abstract}
\begin{multicols}{2}

\section{Introduction}
\label{bevezetes}
Single bubble sonoluminescence (SBSL) is a physical process where an oscillating gas bubble levitated acoustically in a host liquid emits brief flashes of light in each period of the harmonic excitation\ \cite{osszefoglalok}. In experiments the bubble levitation is done in resonators size of a jar, filled with liquid, and the acoustic excitation is accomplished by piezo-electric (PZT) transmitters\ \cite{kiserlet}. The study of SBSL is motivated by the fact that through it one can study physical processes and the properties of matter at extreme conditions of high density, pressure and temperature. The variety of processes associated to the phenomenon include heat\ \cite{hocsere} and mass\ \cite{diffuzio} transfer, chemical reactions\ \cite{kemia}, shape oscillations\ \cite{alakinst}, chaos\ \cite{kaos}, shock-wave\ \cite{lokhull} and light\ \cite{feny} emission to name a few.

In order to validate theories of SBSL it is crucial to be able to measure several important parameters, such as the pressure amplitude near the bubble $P_a$, the ambient radius $R_0$, or the radius-time curve $R(t)$ in an acoustic period. Previous methods to measure these quantities are variations of Mie scattering technique\ \cite{mievariaciok}, direct imaging of the bubble\ \cite{apfel} and a method based on Doppler effect\ \cite{dopler}. The common factor in these methods is that each of them requires a rather sophisticated and expensive experimental setup including lasers, precision optics, and high-speed cameras. Moreover each of these methods are \emph{invasive}, i.e. they involve external action in the measuring process (e.g. laser or back-lighting), which may disturb the measurement of some other features of SBSL. The quantity $P_a$ can also be measured directly by a needle hydrophone, however, because of the high cost and relatively low precision of this, $P_a$ is usually deduced from fitting measured $R(t)$ curves to the solutions of the Rayleigh-Plesset equation which describes the dynamics of the bubble's volumetric oscillations.

These methods confirmed several theoretically predicted features of SBSL, for example that stable light emitting bubbles obey diffusive and chemical stability \cite{HoGa96,GaHo99,KeAp98}. In a previous study \cite{simonPRE} we exploited this fact to develop a new technique to deduce $P_a$, $R_0$ or $R(t)$. Here we elaborate on our method in more detail, present its advantages and limitations and compare its results with Mie scattering.

\section{Description of the fitting technique}
\label{fitting} 
Our technique is based on the following assumptions:

\emph{1. assumption} The dynamics of the bubble wall $R(t)$ is well described by the Rayleigh-Plesset (RP) equation
\begin{eqnarray}
\label{RPeq}
\lefteqn{R\ddot{R}+\frac{3}{2}\dot{R}^{2}=\frac{1}{\rho}[P_{g}(R(t))-P_{f}(t)-P_{0}+P_{v}]}\nonumber\\
&+&\frac{R}{\rho c}\frac{d}{dt}[P_{g}(R(t))-P_{f}(t)]-4\nu\frac{\dot{R}}{R}-\frac{2\sigma}{\rho R},
\end{eqnarray}
where $P_{g}$ is the uniform gas pressure inside the bubble, $P_{f}=-P_{a}\sin(\omega t)$ is the forcing pressure with angular frequency $\omega$, $P_{0}$ is the ambient pressure valid during the measurements, and the remaining parameters are material constants of the host liquid, e.g. $c$ is the speed of sound, $\rho$ its density, and $\nu$ is the kinematic viscosity.
The gas pressure $P_g$ can be related to R(t) through an equation of state. We use a polytropic van der Waals equation of state, modified to include the effects of surface tension $\sigma$ and of vapor pressure $P_v$,
\begin{equation}
\label{VW}
P_{g}(R(t))=\left(P_{0}+\frac{2\sigma}{R_{0}}-P_{v}\right)\frac{(R_{0}^{3}-a^{3})^{\gamma}}{(R(t)^{3}-a^{3})^{\gamma}}.
\end{equation}
Here $a$ is the hard core van der Waals radius of the gas (for argon $a=R_0/8.86$), and $\gamma$ is the ratio of specific heats. In most of the acoustic period except the final stages of the collapse and after-bounces the gas can be considered isothermic \cite{hocsere}, thus we use $\gamma=1$. 

\emph{2. assumption}
Bubbles emitting light in a stable fashion contain only inert gases, and are in stable diffusive equilibrium with the surrounding liquid \cite{indok1}. The points of diffusive equilibrium in the ($P_a, R_0$) space can be calculated from
\begin{equation}
\label{avarages}
C_i/C_0= \frac{\left< P_g\right>_4}{P_{0}^{*}} , \quad \left<X\right>_i= \frac{\int_{0}^{T_a}R(t)^{i}Xdt}{\int_{0}^{T_a}R(t)^{i}dt}.
\end{equation}
where $P_{0}^{*}= 1$\ atm is the standard atmospheric pressure and $T_a$ is the acoustic period. For air in water $C_i$ is the concentration of argon set during the liquid preparation, $C_0$ is the tabulated dissolved air concentration of water under normal conditions. If the liquid preparation is done by degassing, then $C_i$ can be calculated from Henry's law $C_i=0.0093C_0 \cdot P_i/P_{0}^{*}$, where $P_i$ is the partial pressure of air set during the preparation (air contains 0.93\% of argon). The diffusive equilibrium is stable where the curve in the ($P_a, R_0$) space set by equation\ (\ref{avarages}) is characterized by a positive slope.
\emph{3. assumption}
The acoustic pressure amplitude $P_a$ at the bubble's position is directly proportional to the excitation voltage on the piezo transmitters $U_{pzt}$. 
\begin{equation}
\label{fit1}
P_{a}=A\cdot U_{pzt} 
\end{equation}
This assumption is reasonable, when the amplitude of displacement of the resonator walls can be considered small. For values typical in SBSL experiments $P_a < 2$ bar this displacement is on the order of micrometers, thus equation\ (\ref{fit1}) can be used safely. 
We define the dimensionless phase of a flash $\xi=t_{min}/T_a$, as the time elapsed from the beginning of the acoustic period until the bubble reaches its minimum radius $t_{min}$ normalized by the acoustic period. The phase $\xi$ that can be calculated numerically is the sum of the measured phase $\xi_m$ and of a constant parameter $B$, that accounts for  possible phase-shift due to electronics. 
\begin{equation}
\label{fit}
 \xi = \xi_{m}+B,
\end{equation}
The phase can be measured for instance by displaying the PMT signal of the flashes and the sinusoidal excitation voltage $U_{pzt}$ on an oscilloscope.

\begin{figure}[t]
\centering
\epsfig{figure=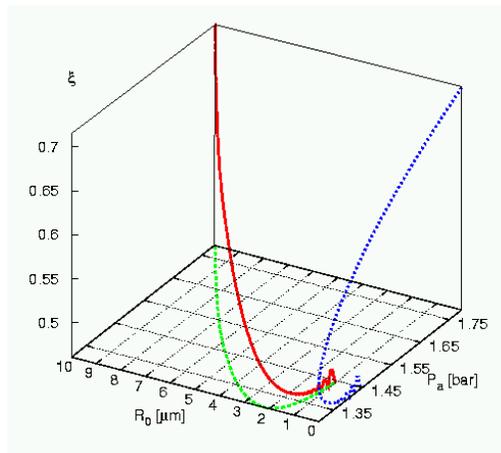,width=6.0cm,angle=270}
\vspace{1cm}
\caption{The curve of diffusive equilibrium in the ($P_a,R_0,\xi$) space, and its projections on the ($P_a,R_0$) and ($P_a,\xi$) planes for $C_i/C_0=0.00135$. Stable sonoluminescing bubbles follow the stable part of this curve (where $dR_0/dP_a >0$, or $d\xi /dP_a > 0$). The advantage of using the quantity $\xi$ is that contrary to $R_0$ it can be easily measured with high precision.}
\label{Fig1}
\end{figure}

The detailed procedure is the following. At an experimentally known dissolved gas concentration, liquid temperature $T$, ambient pressure and driving frequency one measures $\xi_m(U_{pzt})$ at several excitation levels from the smallest possible light emission to the upper limit of SL\ \cite{indok2}. After this equations\ (\ref{RPeq}), (\ref{VW}), (\ref{avarages}) are solved systematically in a wide range of the parameters $P_a$ and $R_0$ and the quantities $\xi$, $\left<P_g\right>_4/P_0^*$ are extracted from the numeric $R(t)$ data. From the resulting $\xi (P_a,R_0,\left<P_g\right>_4/P_0^*)$ data one selects by interpolation a subset for which $\left<P_g\right>_4/P_0^*$ equals the experimentally known value of $C_i/C_0$. This subset is the curve of diffusive equilibrium in the ($P_a,R_0,\xi)$ space (see Fig.\ \ref{Fig1}). The equilibrium is stable where $dR_0/dP_a >0$ or equivalently $d\xi /dP_a > 0$. The final step is to fit the measured $\xi_m(U_{pzt})$ data to the stable part of the diffusive equilibrium curve in the ($P_a,\xi$) plane by adjusting the parameters $A$ and $B$. This fit is highly constrained by the experimental fact that for a given dissolved gas concentration the lower limit of SL is linked to the smallest $P_a$ on the stable diffusive equilibrium curve (see for instance Refs. \cite{KeAp98,HoGa96,GaHo99}). After $A$ and $B$ is found the experimental data points can also be plotted in the ($P_a,R_0$) plane and in principle $R(t)$ curves and other dynamical parameters such as the expansion ratio $R_{max}/R_0$ are also determined. We stress here that only those parts of the numeric $R(t)$ curves contain useful information about the size of the bubble, where the assumption of isothermicity $\gamma=1$ holds. These include the expansion phase, the maximum radius, and the initial stages of the collapse, but exclude the region near the minimum radius and after bounces. 

\section{Experimental Apparatus and the Details of the Measurement}
\label{meres}
\begin{figure}[t]
\centering
\epsfig{figure=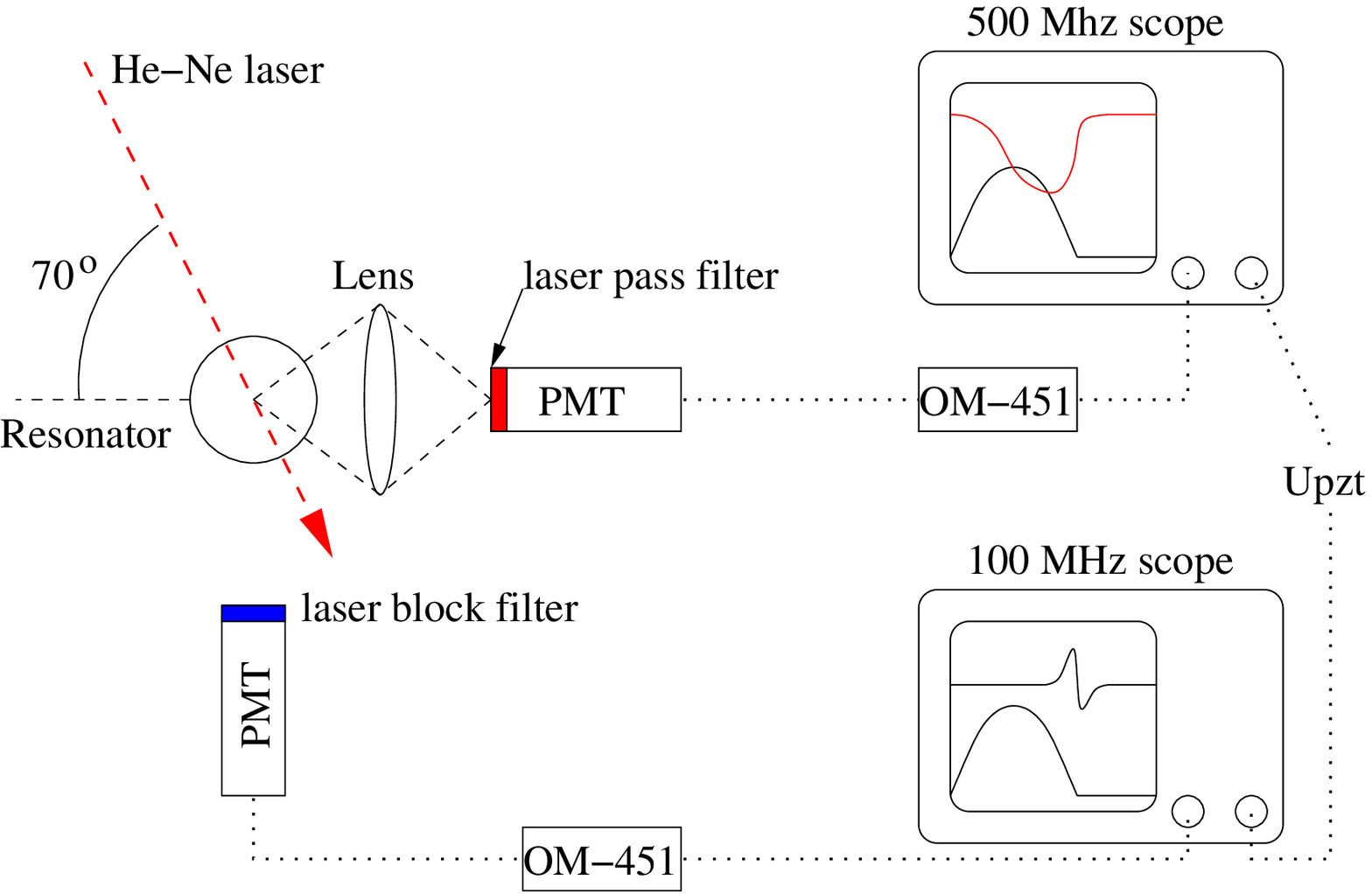,width=8.0cm,angle=0}   
\vspace{1cm}
\caption{The oscillating bubble was illuminated by a 30mW He-Ne laser, and the scattered laser light was focused by a lens through an aperture into the PMT (Hamamatsu R3478). Another PMT of the same kind detected the SL flashes. The signals of the PMTs were amplified and shaped by spectroscopy amplifiers (Ortec-model 451) and visualized together with the monitored electric signal of the piezo transmitters on digital oscilloscopes (HP-54616C 500MHz, HP-54600B 100MHz)}
\label{Fig2}
\end{figure}

To test our method and to compare it with Mie scattering we performed the following experiment. In a degassing equipment \cite{Doc} a distilled water sample was prepared with a dissolved air concentration $0.145C_0$ at a temperature of $T=22^oC$. This corresponds to an argon concentration of $C_i= 0.00135C_0$. The water was transfered to the resonator \cite{Doc} using gravity flow. During the filling the gas pressure in the resonator was kept at the same level as in the degassing equipment, thus no air could diffuse in or out of the water. Such a care had to be taken because the precision of our method is guaranteed only if the gas concentration in the resonator is the same as set by the water preparation. After this the resonator was placed in a setup (Fig.\ \ref{Fig2}) used for recording the SL flashes and the intensities of Mie scattered laser light.

The digital oscilloscopes were triggered by the monitored driving signal $U_{pzt}$. For each excitation level $54$ snapshots of $U_{pzt}$, scattered laser intensity and SL flashes were averaged, and the resulting traces were transfered to a PC. The measurement was done at an excitation frequency of $22700$\ Hz, the water temperature was $T=21.9-22.4^oC$, and the external pressure  $P_0=1017$\ mbar. Figure\ \ref{Fig3}(a) shows the fitting of the measured $\xi(U_{pzt})$ data to the calculated curve of stable diffusive equilibrium at $C_i/C_0=0.00135$. The error of $\xi$ is determined by the jitter in the phase of the flashes $\approx 0.2$\ $\mu s$, thus $\Delta \xi \approx 0.00227$ (the time resolution was 25\ ns), while the error of the pressure amplitude $\Delta P_a = 0.01463$\ bar is calculated from \begin{equation}
\label{Paerror}
\Delta P_a = A\cdot 2\Delta S.
\end{equation}
The quantity $A$ is the proportionality constant from equation\ (\ref{fit1}), and $\Delta S = 0.001563 V$ is the vertical resolution of the digital scope corresponding to the 8-bit digitization. Both the zero-line and the maximum of the $U_{pzt}$ signal are known with a precision of $\Delta S$, thus there is a factor of two in equation\ (\ref{Paerror}).
Using the calculated $\xi(P_a, R_0)$ data and the values of $\xi$ and $P_a$ from the fit of Fig.\ \ref{Fig3}(a) one can plot the measured data points in the ($P_a,R_0$) plane (see the filled squares with error bars in Fig.\ \ref{Fig3}(b)). The errors of $\xi$ and $P_a$ determine a range of $R_0$ in the $\xi(P_a, R_0)$ data, which sets the error bar $\Delta R_0$.


\section{Comparison with Mie scattering}
\label{osszehasonlitas}
The Mie scattering method is based on the detection of laser light scattered from the bubble, whose intensity under certain conditions is proportional to the square of the radius. More precisely the following relation holds
\begin{equation}
\label{Mie}
R(t)^2 =\alpha(U(t)-U_{bg}),
\end{equation}
where $U(t)$ is the output signal of the PMT detecting the scattered intensity, and $U_{bg}$ is the background scattered intensity in the absence of the bubble.
The square-root of the recorded intensity is then fitted to a solution of the RP equation, and the desired parameters $P_a$, $R_0$, $R_{max}$, etc. are determined from the best fit. There are several difficulties of the technique that one must overcome. 

\emph{1.} The scattered intensity has a very strong angle-dependence. This is reduced by choosing an appropriate angle and by use of a lens that averages out the light from different angles. In our measurement we used a setup similar to that of Barber \emph{et al}(1992) in \cite{mievariaciok}. 

\emph{2.} The background scattered intensity is changing on a timescale of seconds because of small dust particles passing by near the position of the bubble due to a slow convection in the resonator. This can be handled partly by using an aperture and a laser beam as narrow as possible, but can not be eliminated completely.

\emph{3.} Modification of the excitation level makes the bubble change its equilibrium position above the pressure anti-node as the averaged Bjerkness and buoyancy forces change\ \cite{Levi}. The intensity of the laser light in the cross-section of the beam is a Gaussian, thus one always has to adjust the direction of the laser, in order to keep the bubble in the center of the beam. For this reason the proportionality constant $\alpha$ in equation\ (\ref{Mie}) may be different for each excitation level depending on how precisely one can follow the bubble. Because of this difficulty we fitted the shape of the measured intensity signals for each excitation level, rather than using a proportionality constant found from a single fit.

We used a numeric code that accomplished the fitting automatically in a user defined range of $P_a$ and $R_0$. The measured $U(t)$ curve was transformed into a normalized positive signal $u(t)= (U(t)-U_{bg})/(U^*-U_{bg})$, where $U^*$ is the minimum of the negative $U(t)$ which corresponds to the maximum radius of the bubble. Then a solution of the RP equation is generated at a given $P_a$ and $R_0$, and the values of $R_{max}$ and $t_{min}$ are extracted from the numeric radius-time curve. After this $R(t)$ was transformed into $r(t) =R(t-t_{min}+t^u_{min})/R_{max}$, where $t^u_{min}$ is the time value of the end of the collapse in the measured $U(t)$ data. This way we got two normalized time series for which the phases corresponding to the minimum radius were identical. To be able to compare $u(t)$ and $r(t)$ these time series had to have the same number of elements, thus we interpolated $r(t)$ at time values taken from $u(t)$. Finally the difference of the signals could be characterized by a single error parameter 
\begin{equation}
\label{error}
\Delta = \frac{1}{N_{min}}  \sum_{i=1}^{N_{min}}| r_i^2 - u_i|,
\end{equation}
where $N_{min}$ is the index of the minimum radius. The reason for calculating the error only until the end of the collapse is that in the after-bounce region the assumption of $\gamma=1$ used in equation\ (\ref{VW}) is not valid anymore. By this strategy for every measured $U(t)$ curve we obtained the values of $\Delta(P_a,R_0)$ in a wide range of $P_a$ and $R_0$ and thus the dynamical parameters appropriate for the smallest error could be identified. Figure\ \ref{cplots} shows the contour-plots of $\Delta$ for a particular case at two levels of the background. As can be seen the fits are best along a line in the ($P_a,R_0$) space, and the fitting is very sensitive to the value of $U_{bg}$.

\begin{figure}[ht]
\centering
\mbox{\subfigure[]{\epsfig{figure=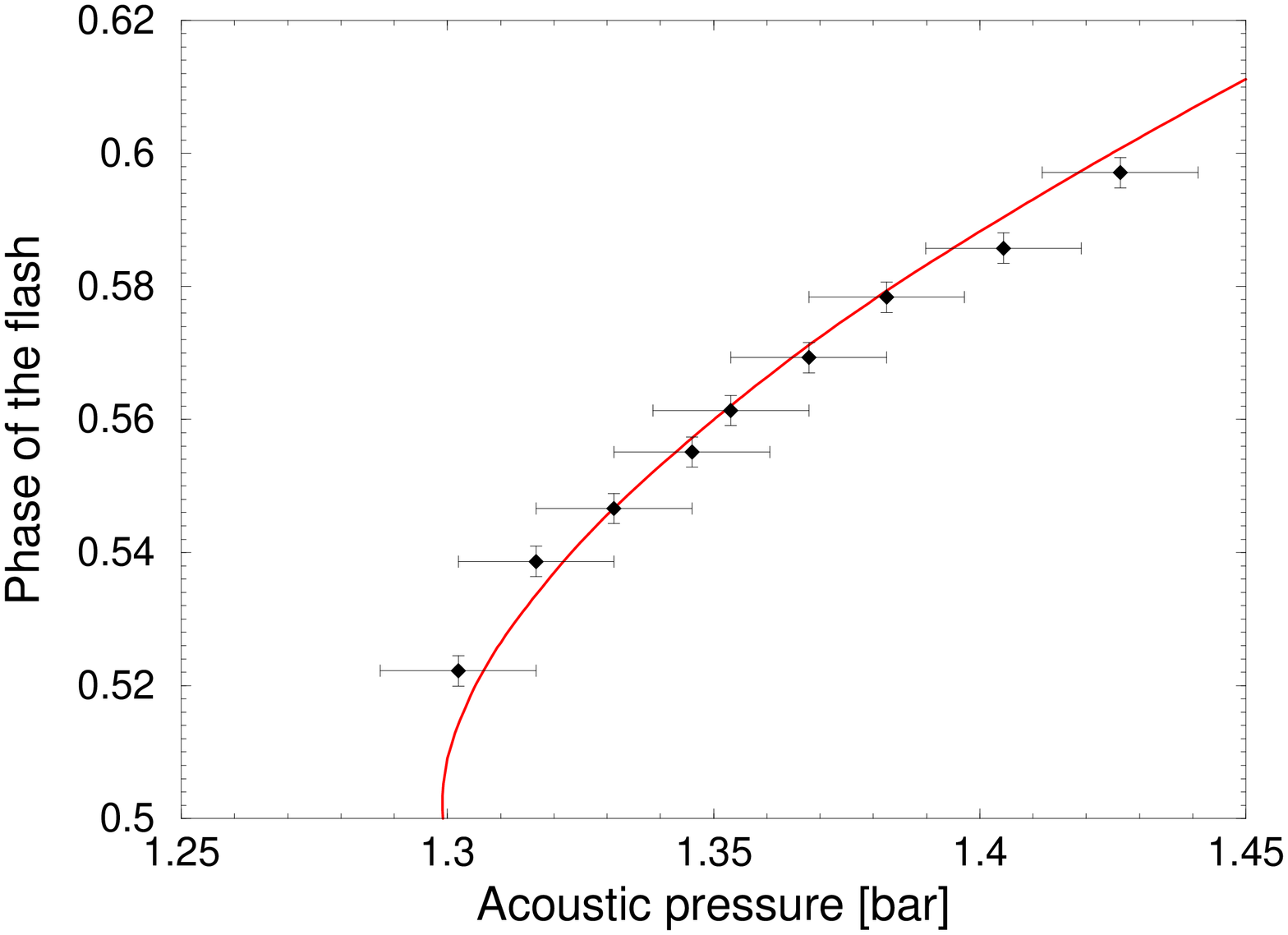,width=6.0cm,angle=270}}
\hspace{1cm}\subfigure[]{\epsfig{figure=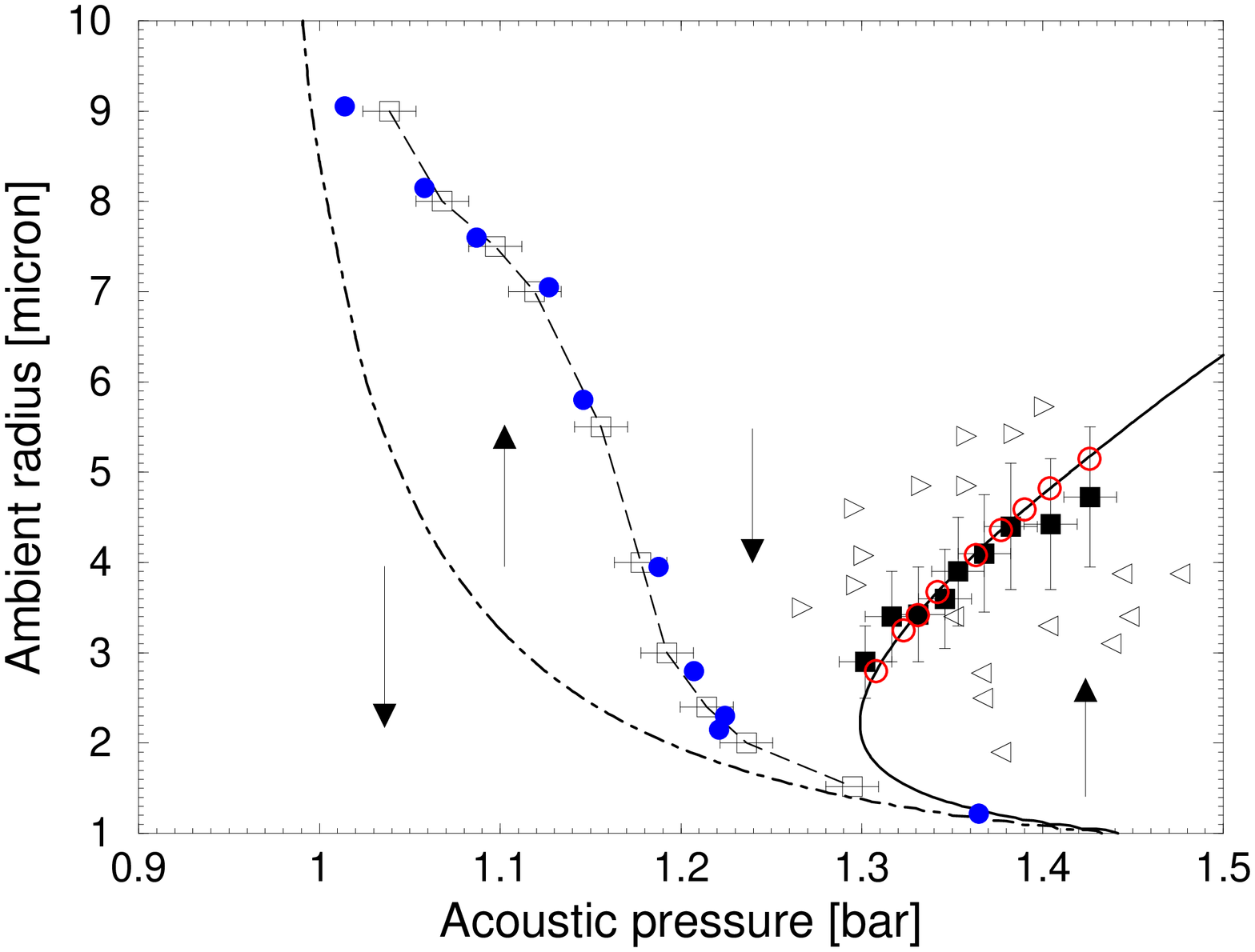,width=6.0cm,angle=270}}}
\caption{The fitting of the measured phases of the flashes to the numerically calculated curve (a), and the same data (filled squares with error bars) in the ($Pa$,$R_0$) plane (b). Also in (b) the left and right triangles correspond to best fits using Mie scattering and the bounding values for the background signal. The  open circles stand for the best Mie fits assuming that equation\ (\ref{avarages}) holds precisely, while the background was allowed to vary between the bounding values. The open squares indicate best Mie fits of non-light-emitting bubbles at $P_a$ values found from the new technique, and finally the filled circles are best Mie fits where both $P_a$ and $R_0$ were fitted. The arrows indicate regions of shrinking and growing bubbles.}
\label{Fig3}
\end{figure}

\begin{figure}[H]
\centering
\mbox{\subfigure[]{\epsfig{figure=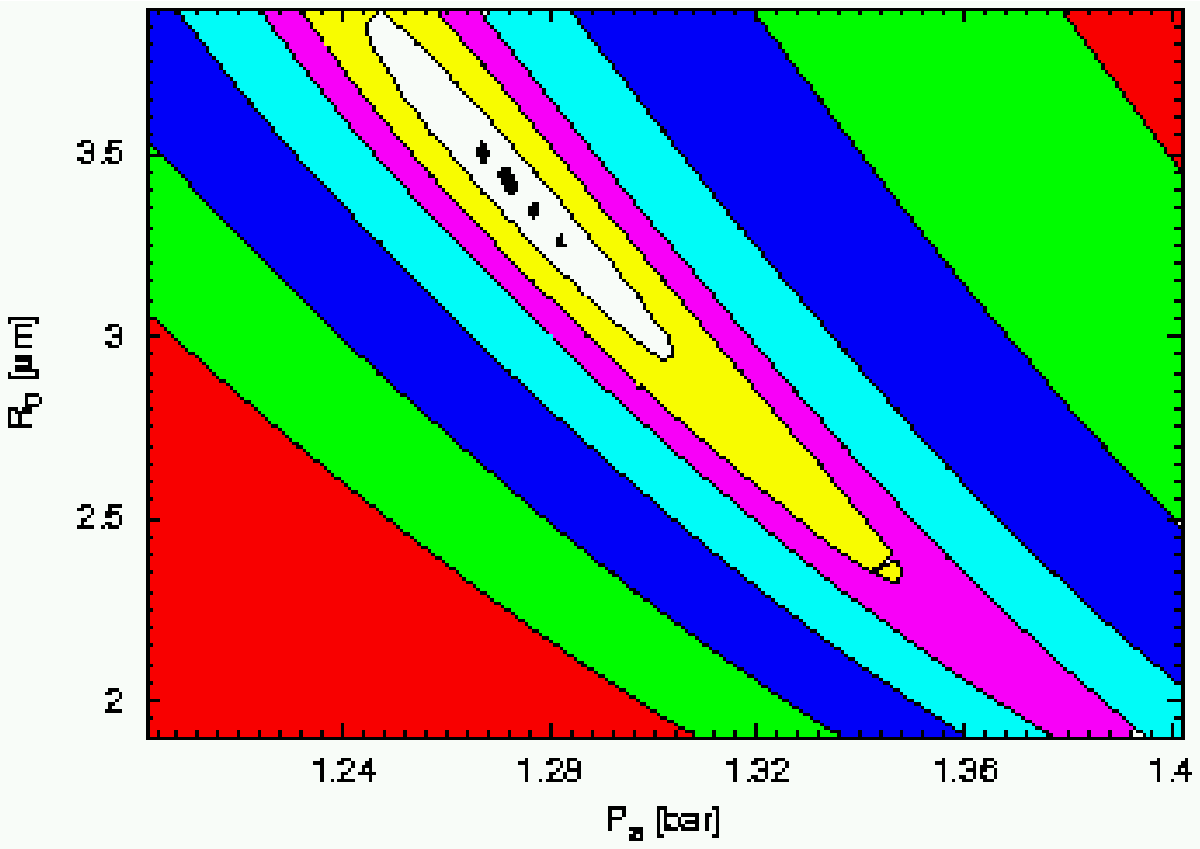,height=8.0cm,angle=270}}
\subfigure[]{\epsfig{figure=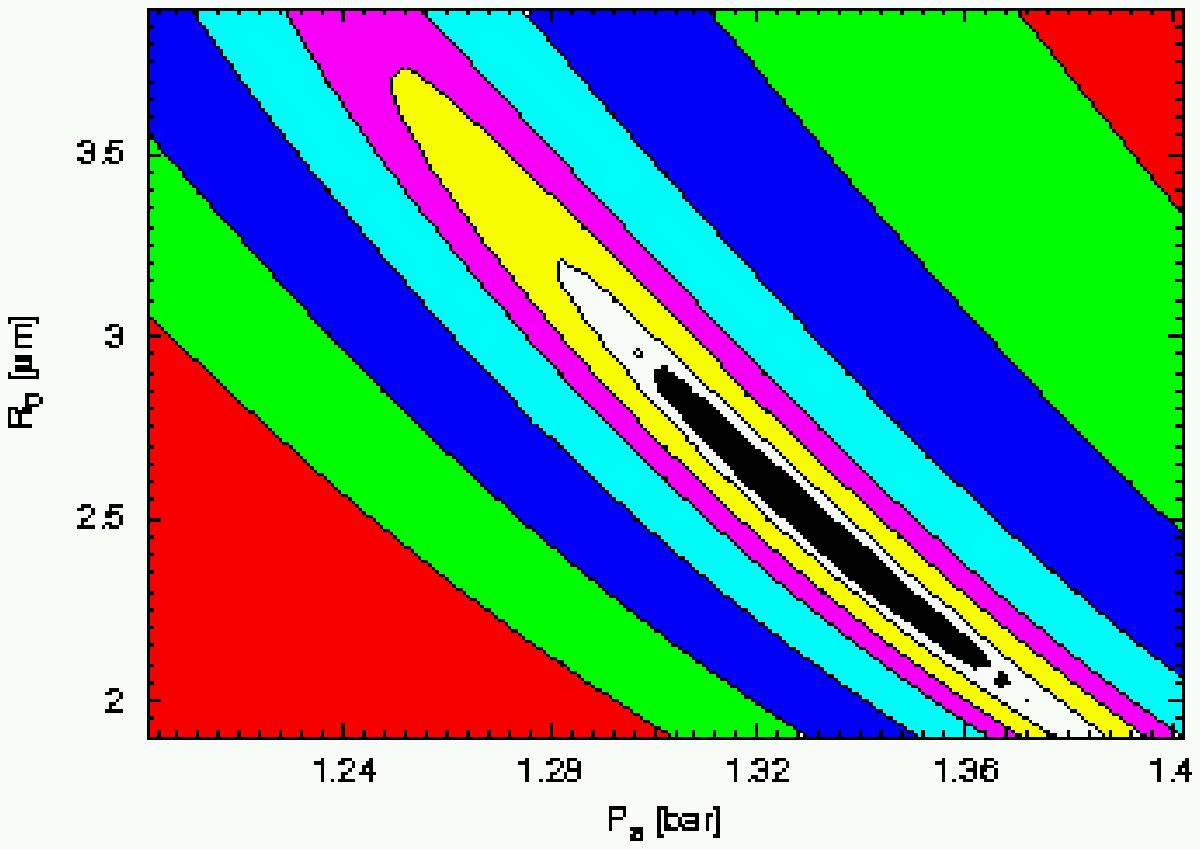,height=8.0cm,angle=270}}
\subfigure{\epsfig{figure=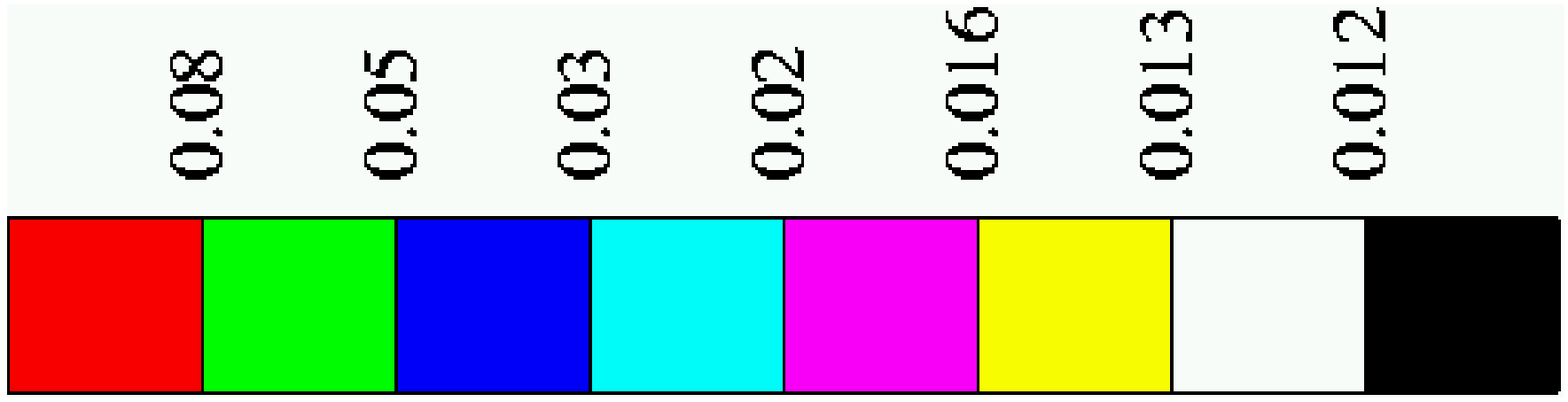,width=5.0cm,angle=270}}}
\caption{The contour-plots of the error of the fit $\Delta$ for the light-emitting bubble at the smallest excitation. The background level used in the fitting was $U_{bg}=-0.0011$\ V in (a), and $U_{bg}=-0.00175$\ V in (b). The average noise level on the intensity signal $U(t)$ corresponds to an error value $\Delta = 0.013$. Fits with $\Delta \leq 0.013$, all look good to the eye.}
\label{cplots}
\end{figure}

In our measurements the precision of $U_{bg}$ was constrained by the vertical resolution of the digital scope and the previously mentioned slow variance due to moving dust particles to $-0.0011 \geq U_{bg} \geq -0.00225$ (Volts). For comparison the values of $U^*$ corresponding to the maximum radii were $-0.05317..-0.07742$ (Volts). In Fig.\ \ref{cplots} the points within the contour $\Delta = 0.013$ (which equals the noise level on the scattered intensity) determine the errors of $\Delta P_a \sim 0.04$\ bar and $\Delta R_0 \sim 0.7$\ $\mu m$. If we also includes the uncertainty of $U_{bg}$ then the final estimates are $\Delta P_a \sim 0.08$\ bar and $\Delta R_0 \sim 1$\ $\mu m$. These values are consistent with the errors given by others \cite{KeAp98,GaHo99}. 
 The best fits with the bounding values of the background are shown in Fig.\ \ref{Fig3}(b), where in the case of left and right triangles the background was $U_{bg}=-0.0011$, and $U_{bg}=-0.00225$ accordingly. The filled squares found from the fitting technique of sec.\ \ref{fitting} in each case lie between the best Mie fits confirming that these two methods provide identical results within the experimental uncertainties. 

\begin{figure}[t]
\centering
\mbox{\subfigure{\epsfig{figure=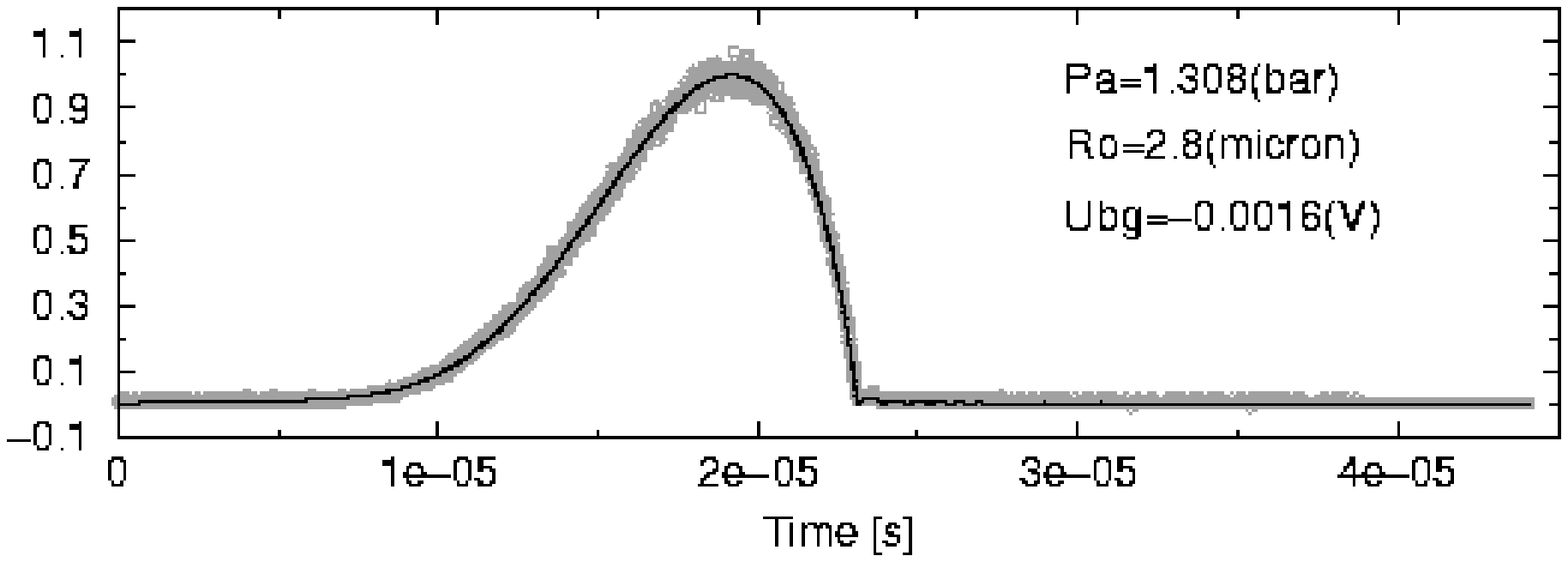,height=8.5cm,angle=270}}
\hspace{0.5cm}\subfigure{\epsfig{figure=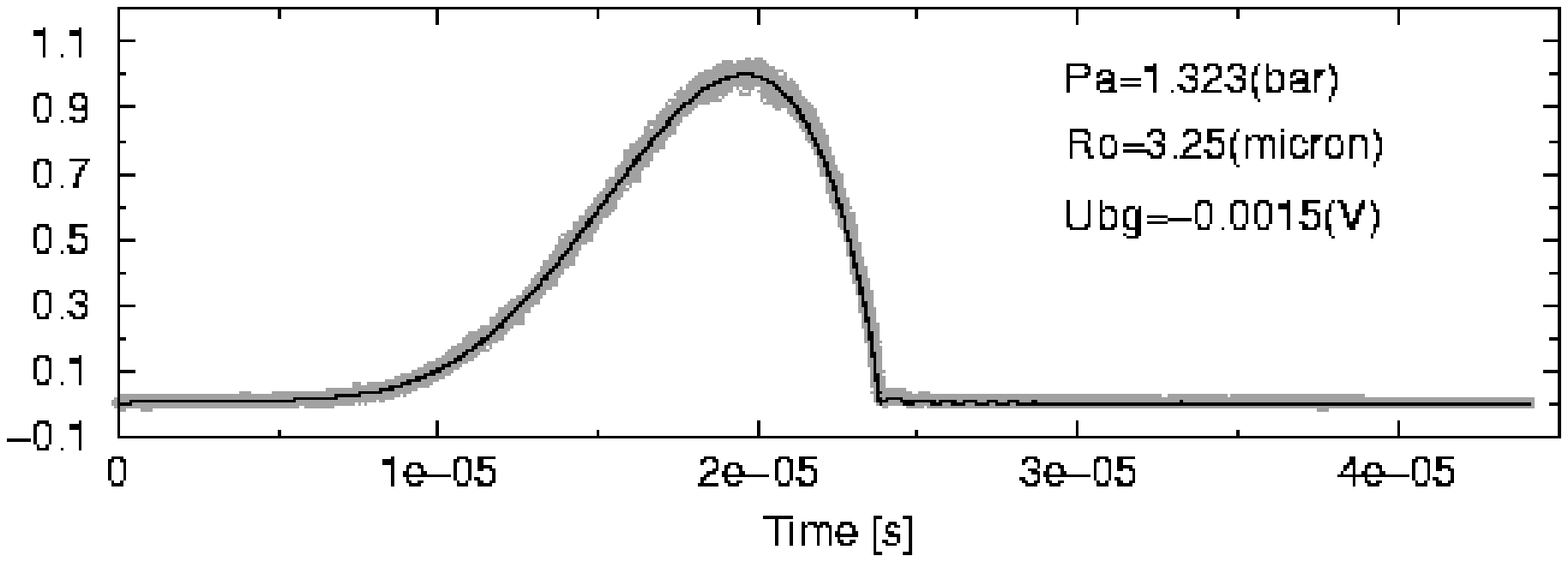,height=8.5cm,angle=270}}}
\mbox{\subfigure{\epsfig{figure=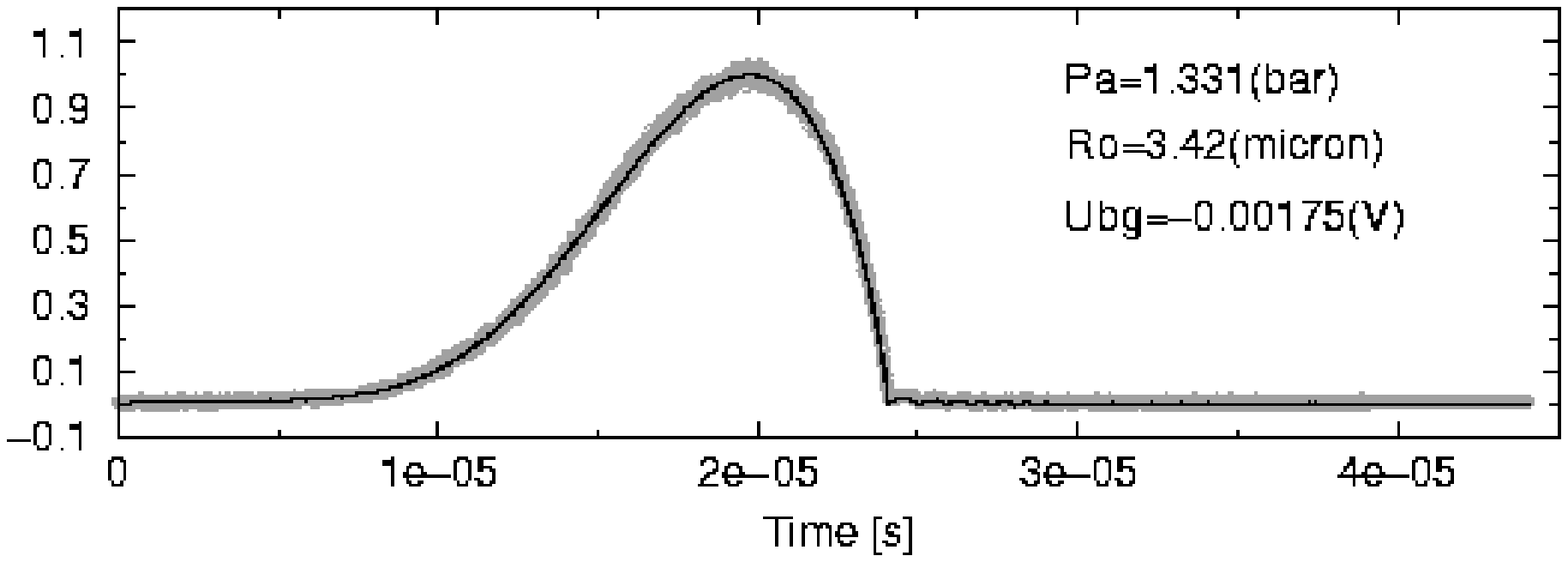,height=8.5cm,angle=270}}
\hspace{0.5cm}\subfigure{\epsfig{figure=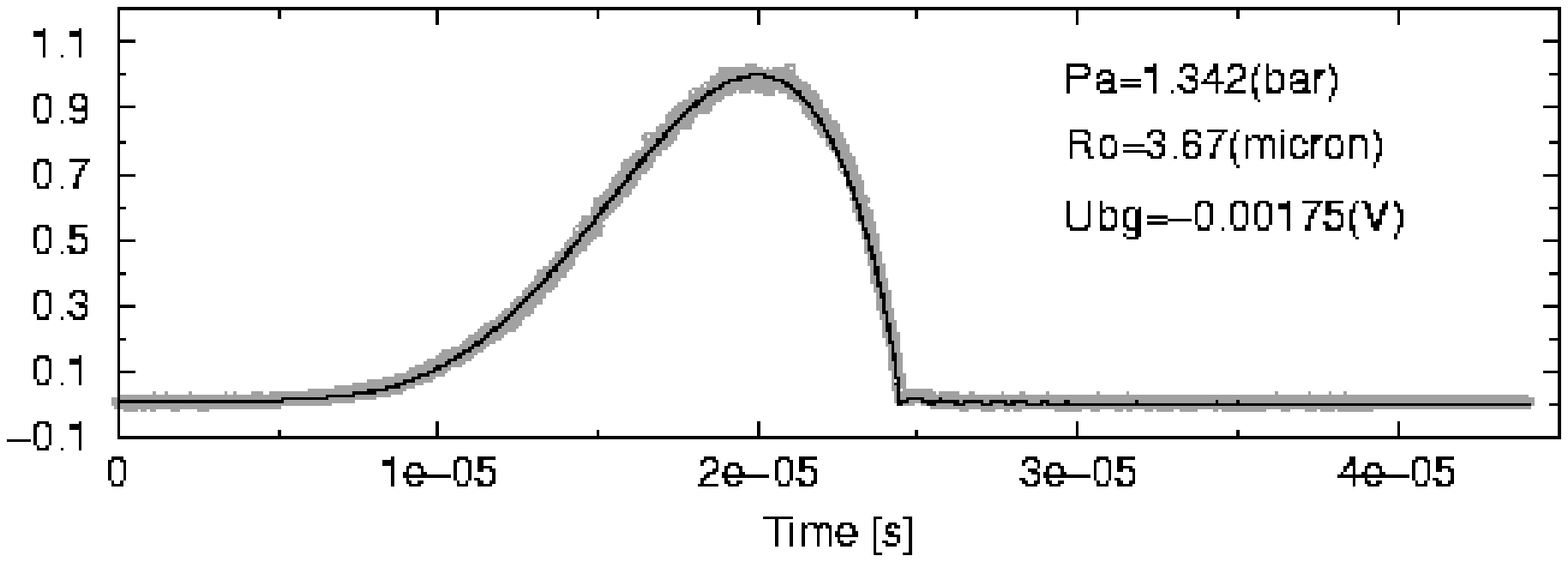,height=8.5cm,angle=270}}}
\mbox{\subfigure{\epsfig{figure=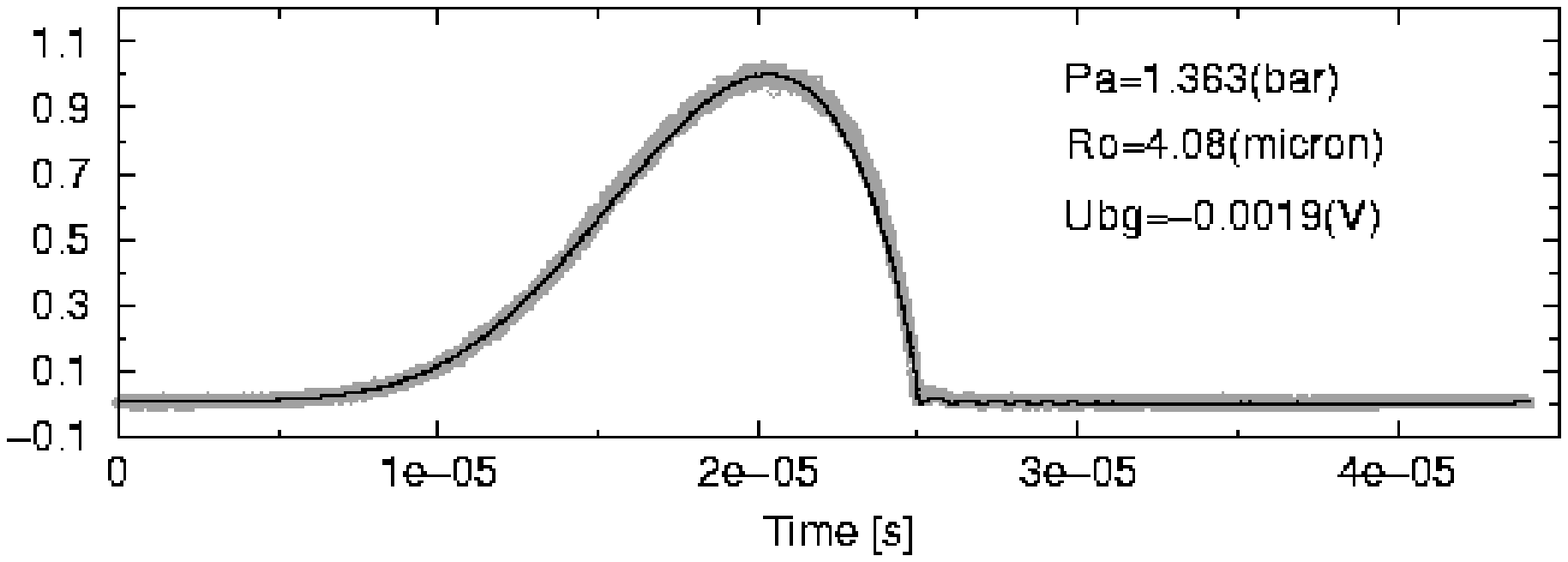,height=8.5cm,angle=270}}
\hspace{0.5cm}\subfigure{\epsfig{figure=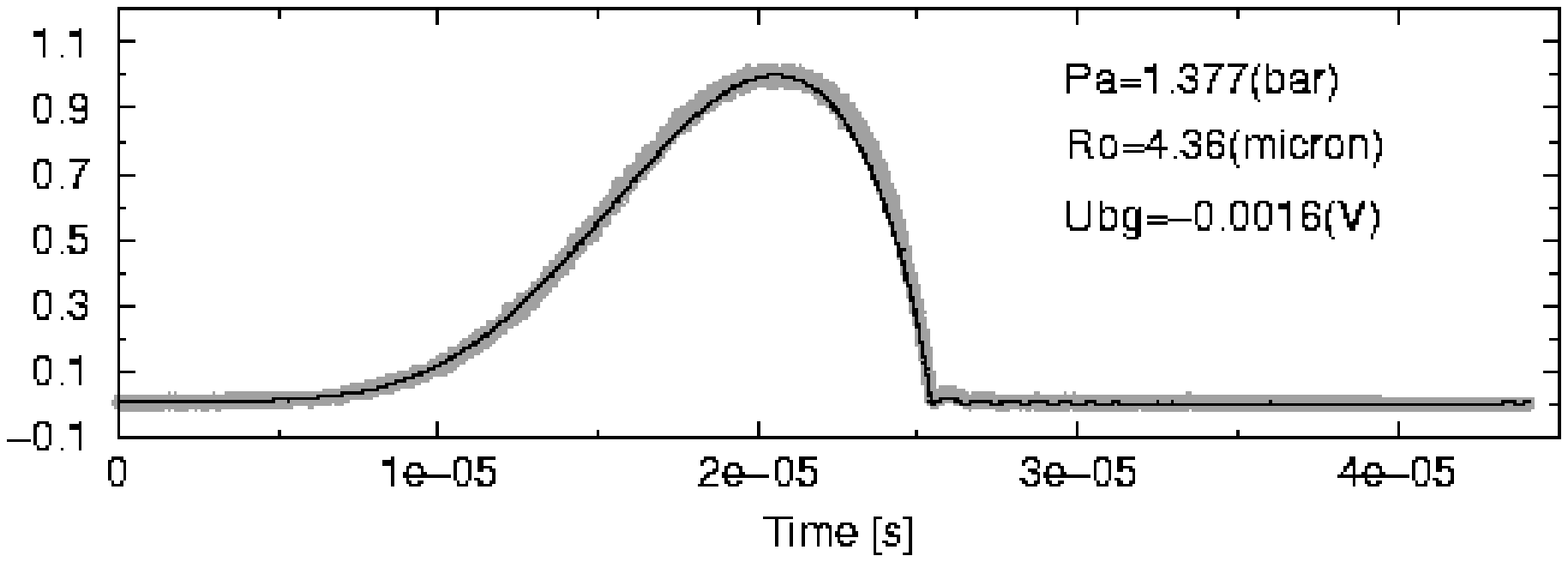,height=8.5cm,angle=270}}}
\mbox{\subfigure{\epsfig{figure=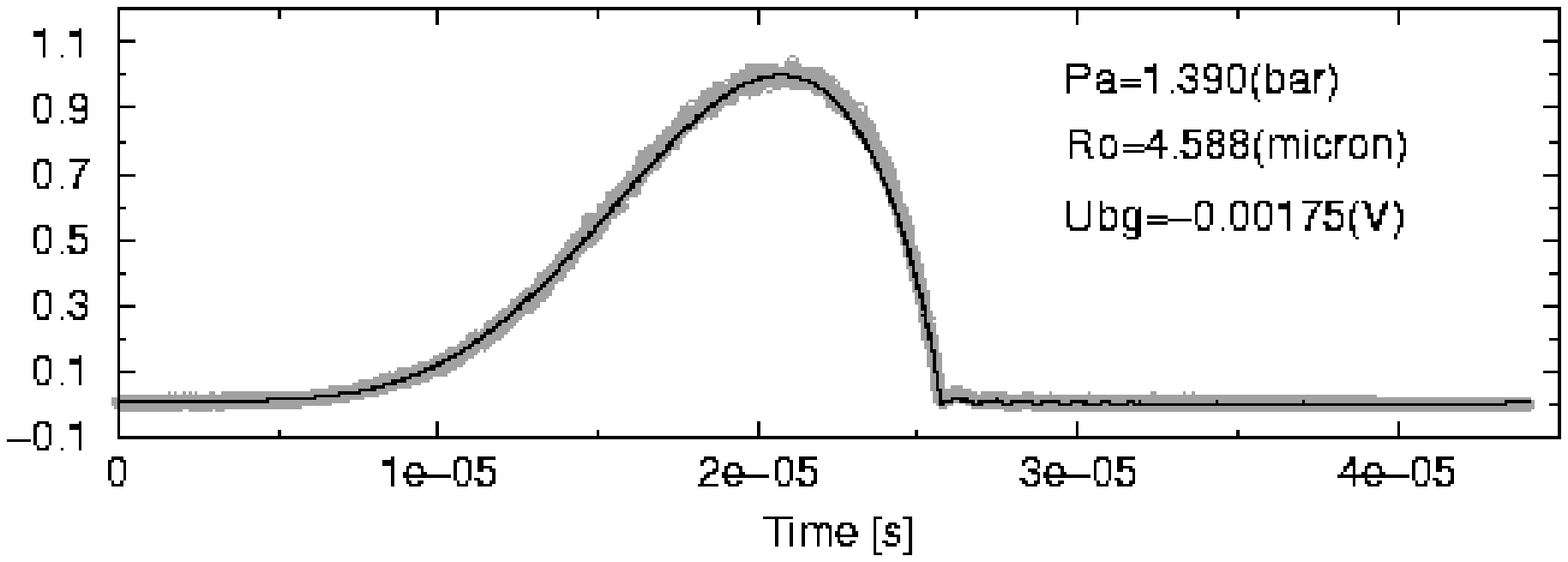,height=8.5cm,angle=270}}
\hspace{0.5cm}\subfigure{\epsfig{figure=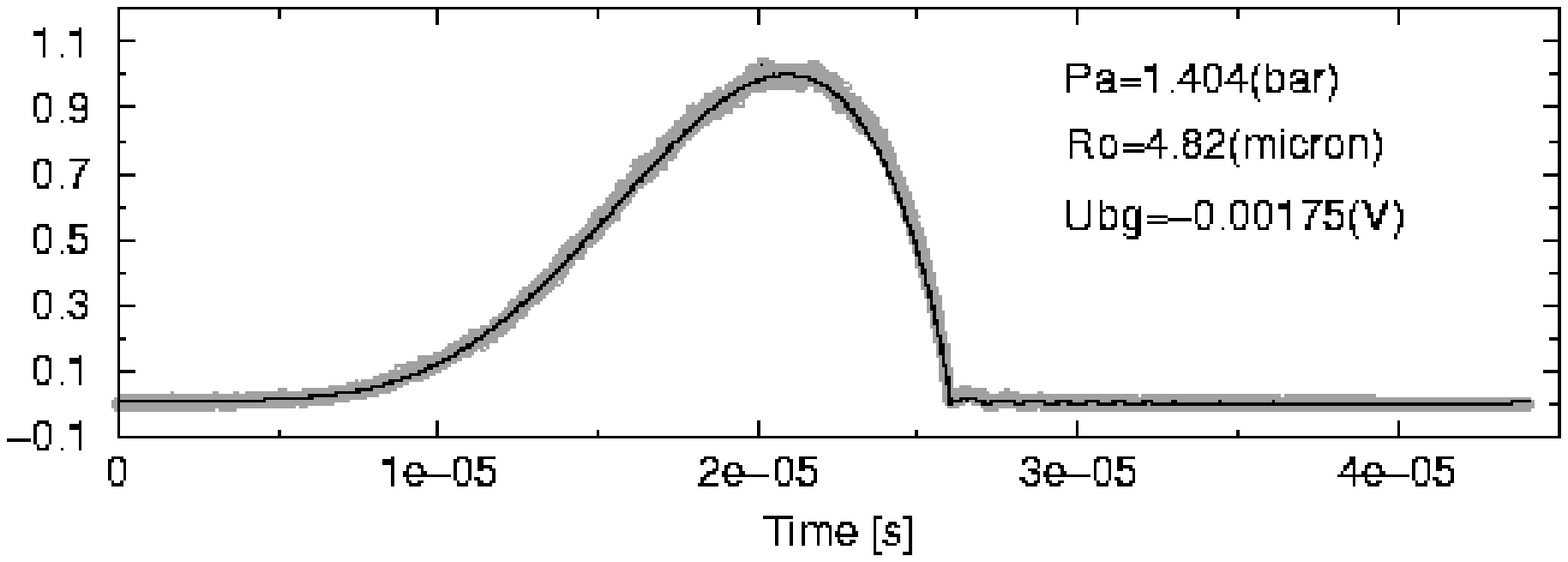,height=8.5cm,angle=270}}}
\mbox{\subfigure{\epsfig{figure=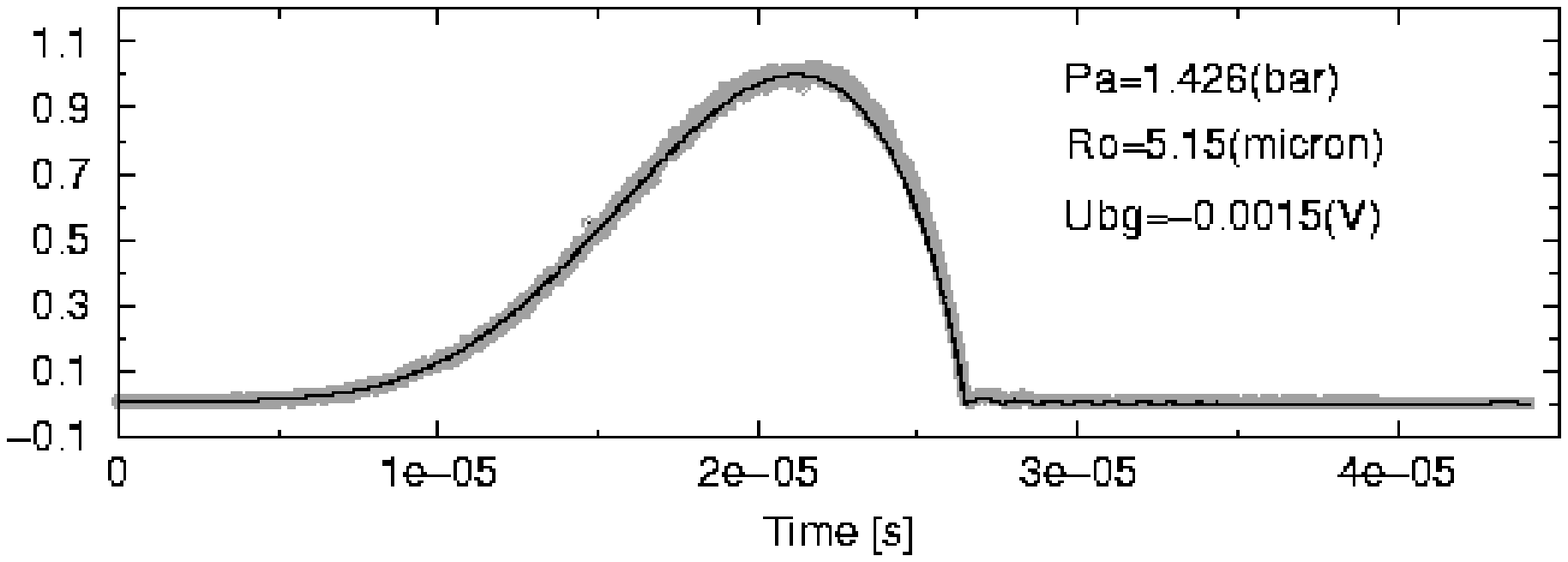,height=8.5cm,angle=270}}}
\caption{The best fits  of the normalized $u(t)$ and $r(t)^2$ time series corresponding to the open circles in Fig.\ \ref{Fig3}(b). The parameters of the fit are indicated in the corner of the figures.}
\label{Miefittelesek}
\end{figure}

For further confirmation of the assumptions in sec.\ \ref{fitting} we also fitted the Mie scattered data by a different strategy, where the background was allowed to vary between $-0.0011$ and $-0.00225$, and the $R(t)$ data of the fits had to satisfy equation\ (\ref{avarages}) exactly. The best fits of this strategy are shown as open circles in Fig.\ \ref{Fig3}. 
As can be seen all the circles are within the error-bars found from the new technique. The quality of these Mie-fits are also shown individually in Figs.\ \ref{Miefittelesek}. 

For another independent test to confirm the assumptions of the new technique we also analyzed measurements of $U_{pzt}$ and Mie scattered data of non-light-emitting "bouncing" bubbles. The open squares with error bars in Fig.\ \ref{Fig3}(b) show the case where the pressure amplitudes were calculated from equation\ (\ref{fit1}) using the value of $A$ found from the fitting of the light emitting bubbles. The ambient radii were determined afterward by best Mie fits at the prescribed $P_a$ and by using the average $U_{bg}$. Then the fitting of the Mie scattered data was also accomplished by using the same $U_{bg}$, but both the $P_a$ and $R_0$ parameters are allowed to vary, and the information of the $U_{pzt}$ data is not used. The best fits using this strategy are (shown as filled circles in Fig.\ \ref{Fig3}(b)) in excellent agreement with the best fits of the previous case.
 The phase diagram of Fig.\ \ref{Fig3}(b) can also be compared to a measurement of reference\ \cite{GaHo99}, where the experimental conditions were quite close to ours (water was prepared with a dissolved air concentration of $0.14C_0$, and the excitation frequency was $20.6$\ KHz. The comparison with Fig.\ 4 of \cite{GaHo99} shows a good quantitative agreement further justifying our method.

\section{Conclusion}
\label{konkluzio}
In this paper we presented an alternative method for the deduction of important dynamical parameters of SBSL, namely the pressure amplitude $P_a$, the ambient radius $R_0$, and the expansion ratio $R_{max}/R_0$. The method is based on previously confirmed theoretical assumptions and the measurement of the phase of the flashes in the acoustic period together with the driving signal on the piezo transmitters, and the precise knowledge of the experimental conditions, especially the dissolved gas concentration in the liquid. We demonstrated the method by measuring the $P_a$ and $R_0$ parameters of light emitting bubbles for a given argon concentration. At the same time measurements of these parameters were also done by the conventional Mie scattering method, revealing that within experimental uncertainties these two approaches produced identical results. The phase diagram obtained is in good quantitative agreement with that of reference\ \cite{GaHo99}. The limitation of the method is that unlike Mie scattering it can not be used to measure non-light-emitting bubbles, and also unable to measure unstable SL. However the important advantage of the method is its \emph{non-invasive} chararcter, and thus the applicability to situations where the other methods are either impractical or even potentially destructive to use as e.g. in single photon correlation measurements, or in measuring the parametric dependence of SBSL spectra. Moreover the technique does not require expensive apparatus other than a digital oscilloscope and a PMT, and the data analysis is easily automatized. 
\section*{Acknowledgements}
\label{koszonet}
The authors acknowledge financial support by the Danish Nonlinear School and the Danish National Science Foundation. G. Simon also thanks the ERASMUS student grant for financial support, and the hospitality of the Niels Bohr Institute where the measurements were done. Attila Simon is thanked for sleeping. 

\end{multicols}

\newpage

\begin{thebibliography}{99}
\bibitem{osszefoglalok}
B. P. Barber \emph{et al}, Phys. Rep. 281, 65 (1997); M. P. Brenner, S. Hilgenfeldt and D. Lohse, Rev. Mod. Phys, 74, 425 (2002).
\bibitem{kiserlet}
D. F. Gaitan, Ph.D. thesis, University of Mississippi, 1990; S. J. Putterman, Sci. Am. 272, 32 (1995).
\bibitem{hocsere}
A. Prosperetti, J. Acoust. Soc. Am. 61, 17 (1977); A. Prosperetti, J. Fluid Mech. 222, 587 (1991); K. Yasui, J. Acoust. Soc. Am. 98, 2772 (1995), M. -C. Chu, and D. Leung, J. Phys Condens. Matter 9, 3387 (1997);
\bibitem{diffuzio}
A. Eller and H. G. Flynn, J. Acoust. Soc. Am. 37, 493 (1965); M. M. Fyrillas and A. J. Szeri, J. Fluid Mech. 277,381 (1994); R. L\"ofstedt, K. R. Weninger, S. J. Putterman, B. P. Barber, Phys. Rev. E 51, 4400 (1995); M. P. Brenner, D. Lohse, D. Oxtoby, and T. F. Dupont, Phys Rev. Lett. 76, 1158 (1996); P. H. Roberts, C. C. Wu, J. Theoret. Comput. Fluid Dynamics, 10, 357 (1998)
\bibitem{kemia}
D. Lohse, M. Brenner, T. Dupont, S. Hilgenfeldt, and B. Johnston, Phys. Rev. Lett. 78, 1359 (1997); D. Lohse, S. Hilgenfeldt, J. Chem. Phys. 107, 6986 (1997); K. Yasui, J. Phys. Soc. Jpn. 66, 2911 (1997).
\bibitem{alakinst}
A. Prosperetti, Quart. Appl. Math. 34, 339 (1977); M.P. Brenner, D. Lohse, and T.F. Dupont, Phys Rev. Lett. 75, 954 (1995); S. Hilgenfeldt, D. Lohse, and M.P. Brenner, Phys. Fluids 8, 2808 (1996); C.C. Wu, P.H. Roberts, Phys. Lett. A, 250, 131 (1998); A. Prosperetti and Y. Hao, Philos. Trans. R. Soc. London A357, 203 (1999). U. H. Augsd\"orfer, A. K. Evans, D. P. Oxley, Phys. Rev. E 61, 5278 (2000); B. D. Storey, Phys. Rev. E 64, 017301.
\bibitem{kaos}
W. Lauterborn and E. Suchla, Phys. Rev. Lett. 53, 2304 (1984); W. Lauterborn and A. Koch, Phys. Rev. A 35, 1974 (1987); P. Smereka, B. Birnir, and S. Banerjee, Phys. Fluids 30, 3342 (1987); U. Parlitz, V. Englisch, C. Scheffczyk, and W. Lauterborn, J. Acoust. Soc. Am. 88, 1061 (1990); R. G. Holt, D. F. Gaitan, and A. A. Atchley, Phys. Rev. Lett. 72, 1376 (1994); G. Simon \emph{et al}, Nonlinearity 15, 25 (2002).  
\bibitem{lokhull}
T. J. Matula, I. M. Hallaj, R. O. Cleveland, and L. A. Crum, J. Acoust.Soc. Am. 103, 1377 (1997); J. Holzfuss, M. R\"{u}ggeberg, and A. Billo, Phys. Rev. Lett. 81, 5434 (1998); R. Pecha and B. Gompf, Phys. Rev. Lett. 84, 1328 (2000).
\bibitem{feny}
B. P. Barber and S. J. Putterman, Nature (London) 352, 318 (1991); R. Pecha, B. Gompf, G. Nick, Z.Q. Wang, and W. Eisenmenger, Phys. Rev. Lett. 81, 717 (1998); B. P. Barber, C. C. Wu, R. L\"ofstedt, P. H. Roberts, and S. J. Putterman, Phys. Rev. Lett. 72, 1380 (1994); R. A. Hiller, S. J. Putterman, and B. P. Barber, Phys. Rev. Lett. 69, 1182 (1992); W. C. Moss \emph{et al}, Phys. Rev. E 59, 2986 (1999); S. Hilgenfeldt, S. Grossmann, and D. Lohse, Phys. Fluids 11, 1318 (1999).
\bibitem{mievariaciok}
B. P. Barber, S. J. Putterman, Phys. Rev. Lett. 69, 3839 (1992); W. J. Lentz, A. A. Atchley, and D. F. Gaitan, Appl. Opt. 34, 2648 (1995); K. R. Weninger, B. P. Barber, and S. J. Putterman, Phys. Rev. Lett. 78, 1799 (1997); K. R. Weninger, P. G. Evans, and S. J. Putterman, Phys. Rev. E 61, 1020 (2000); B. Gompf, and R. Pecha, Phys. Rev. E 61, 5253 (2000);
\bibitem{apfel}
Y. J. Tian, J. A. Ketterling, and R. E. Apfel, J. Acoust. Soc. Am. 100, 3976 (1996).
\bibitem{dopler}
G. A. Delgadino, and F. J. Bonetto, Phys. Rev. E 56, R6248 (1997); G. Vacca, R. D. Morgan, and R. B. Laughlin, Phys. Rev. E 60, 6303 (1999).
\bibitem{HoGa96}
R. G. Holt and D. F.Gaitan, Phys. Rev. Lett. 77, 3791 (1996).
\bibitem{GaHo99}
D. F. Gaitan and R. G. Holt, Phys. Rev. E 59, 5495 (1999).
\bibitem{KeAp98}
J. A. Ketterling and R. E. Apfel, Phys. Rev. Lett. 81, 4991 (1998); J. A. Ketterling and R. E. Apfel, J. Acoust. Soc. Am. 107, L13 (2000); J. A. Ketterling and R. E. Apfel, Phys. Rev. E 61, 3832 (2000).
\bibitem{simonPRE}
G. Simon, I. Csabai, \'A. Horv\'ath, and F. Szalai, Phys. Rev. E 63, 026301.
\bibitem{indok1}
This assumption is the direct consequence of chemical reactions \cite{kemia} and diffusion, and it was confirmed experimentally in \cite{MaCr98,KeAp98,HoGa96,GaHo99}.
\bibitem{indok2}
At high degasing a hysteresis in the lower SL threshold can be observed (see Lohse and Hilgenfeldt in \cite{kemia}). In these cases the state with the smallest light emission should be approached from above to obtain the true values of the lower SL threshold. 
\bibitem{MaCr98}
T. J. Matula and L. A. Crum, Phys. Rev. Lett. 80, 865 (1998).
\bibitem{Doc}
The detailed description of the experimental apparatus and procedures will be available in G. Simon, PhD thesis, E\"otv\"os University (to be published).
\bibitem{Levi}
T. J. Matula, S. M. Cordry, R. A. Roy, and L. A. Crum, J. Acoust. Soc. Am. 102, 1522 (1997).
\end{thebibliography}
\end{document}